\documentclass[onecolumn,amsmath,amssymb,pra,superscriptaddress]{revtex4}

\usepackage{graphics}
\usepackage{epsfig}
\usepackage{amsmath}
\usepackage{amsfonts}
\usepackage{color}
\usepackage{dcolumn}
\usepackage{bm}
\usepackage{mathrsfs}
\usepackage{float}

\begin{document}

  \bibliographystyle{apsrev}
  
\title{Near-Exact Non-relativistic Ionisation Energies for Many-Electron Atoms }

      \author{E. O. Jobunga}
       \affiliation{\it  Department  of Mathematics and Physics , Technical University   of Mombasa,\\ P. O. Box 90420-80100, Mombasa, Kenya}
       

\begin{abstract}
Electron-electron interactions and correlations form the basis of difficulties encountered in the theoretical solution of problems dealing with multi-electron systems. Accurate treatment of the electron-electron problem is likely to unravel some nice physical properties of matter embedded in the interaction. In an effort to tackle this many-body problem, a symmetry-dependent all-electron potential generalised for an $n$-electron atom is suggested in this study. The symmetry dependence in the proposed potential hinges on an empirically determined angular momentum dependent partitioning fraction for the electron-electron interaction. With the potential, all atoms are treated in the same way regardless of whether they are open- or closed-shell using their system specific information. The non-relativistic groundstate ionisation potentials for atoms  with up to $103$ electrons generated using the all-electron potential are in reasonable agreement with the existing experimental and theoretical data. The effects of higher-order non-relativistic interactions as well as the finite nuclear mass of the atoms are also analysed. 
\end{abstract}

\maketitle

\section{Introduction}

The theory of quantum many-body systems is an effective theoretical structure and solvable approach of understanding the collective behaviour of the interacting many-particle systems\cite{Hugenholtz1965}. The solution of the many-electron problem is important because electrons determine the physical properties of materials and molecules. Many-body physics is heavily applicable in condensed matter, Bose-Einstein Condensation (BEC) and superfluidity, quantum chemistry, atomic, molecular, nuclear physics, as well as quantum chromodynamics.

Electron correlation energy, among the interacting many-body particles, is defined as the difference between the exact non-relativistic energy eigenvalue of the electronic Schr\"odinger equation and the energy of the single configuration state function (CSF) approximation, commonly called the Hartree-Fock energy \cite{Verdebout2013}. 

Accurate description of electron-electron interaction remains a major challenge in atomic structure calculations \cite{Verdebout2013}. To meet this challenge, a number of different methods have been developed such as the many-body perturbation theory (MBPT) \cite{Tobocman1957}, configuration interaction (CI)\cite{Cremer2013},  density functional theory \cite{Kohn1965}, coupled cluster theories, and different kinds of variational methods \cite{Cramer2002}. Hylleraas-type calculations \cite{Hylleraas1929} is an example of the variational methods in which the interelectronic distance $r_{12}$ is employed explicitly in the construction of the wavefunction resulting into the most accurate eigenvalues, although computationally expensive.

A pseudopotential, or an optimized potential, is an effective potential used as an approximation for the simplified description of complex atoms, molecules, and other quantum systems. The use of pseudoptentials was first introduced by Fermi \cite{Cohen1984}. Hellmann \cite{Hellmann1935} subsequently developed a pseudopotential model for atoms which has been extensively used in atomic scattering \cite{Callaway1969}. The use of pseudopotential method in the many-body problems is computationally less expensive and has the potential of revealing the underlying processes in the interaction dynamics.

In this work, a central screening potential in an independent particle model introduced in our previous papers \cite{Jobunga2018, Jobunga2017b, Jobunga2017c, Jobunga2020}, based on an alternative multipole expansion of the electron-electron interaction \cite{Jobunga2017e}, is extended to incorporate the expected symmetry dependence of the electron-electron interaction in the Hamiltonian of an $n-$electron atom. The generalised all-electron potential developed in this work is then used to evaluate the groundstate ionisation potentials of atoms with up to $103$ electrons. 

Atomic systems have been chosen to test the validity and efficiency of the present method in predicting experimental data. The atomic systems are not only useful as a playground to for testing physical approximations and numerical algorithms but also provide a basis for understanding complex systems like molecules and condensed matter. Several theoretical studies have been undertaken on atomic systems\cite{Kotochigova1997,Kraisler2010,Kraisler2013,
Klopper2010} with DFT methods widely adopted. Eventhough DFT method with LSDA and GGA functions have been used to generate total and ionisation energies which are in good agreement with experimental results for up to $86$ electrons \cite{Kraisler2010,Kraisler2013}, further improvement in the description of atomic systems focussing on non-spherical calculations and on advanced non-local correlation functional is recommended \cite{Kraisler2010}.  

The symmetry-dependent all-electron potential suggested in this study naturally includes  the non-spherical terms of the multipole series expansion of the electron-electron interaction. In the present method, we achieve total separability of the Hamiltonian of the many-electron atom and hence the calculations can be considered to yield near-exact non-relativistic eigenvalues. Our results are compared with reported literature data and with results generated using our previously developed potential which was based on the classical partitioning of the electron-electron interaction \cite{Jobunga2017c}.

\section{Theory}
The non-relativistic Hamiltonian of an $n$-electron system with a nuclear charge $Z$ is given by

    \begin{equation}
    \mathrm{H} = \sum_i^n \left[\frac{p_i^2}{2} - \frac{Z}{r_i} +  \sum_{j\neq i}^{n-1}\frac{1}{|\mathbf{r}_i-\mathbf{r}_j|} \right]
    \end{equation}
where the first term on the right corresponds to the kinetic energy of the $i^{\mathrm{th}}-$electron, the second term corresponds to the  interaction of the $i^{\mathrm{th}}-$electron with the nuclear charge, and the last term in the summation corresponds to the interaction between the $i^{\mathrm{th}}-$ and $j^{\mathrm{th}}-$ electron. The second and the last term form the potential energy function of a bound $n$-electron system.

 In our previous work \cite{Jobunga2017,Jobunga2017e}, it was shown that the electron correlation interaction analytically simplifies to
 \begin{equation}
  \begin{split}
  \frac{1}{\mid \vec{r}_i-\vec{r}_j \mid} &= (r_i^2 -2r_i\,r_j\,x +r_j^2)^{-1/2} \\
 &=\frac{4 \pi}{\sqrt{r_i^2+r_j^2}}\sum_l \tilde{j}_l(r_i,r_j)\, Y_l^{m*}(\hat{r_i})\, Y_l^{m}(\hat{r_j}),\\  \label{eq:0a}
  \end{split}
  \end{equation}
  
  where ${x= \cos \theta}$, $\theta$ is the angle between vectors $\vec{r}_i$ and $\vec{r}_j$, the unit vector $\hat{r}$ specifies the angular coordinates of vector $\vec{r}$, and  
  \begin{equation}
  \tilde{j}_l(r_i,r_j) = \sum_{k=l,l+2, \cdots} \frac{(2k-1)!!}{(k-l)!!\,(k+l+1)!!}\, \left(\frac{r_ir_j}{r_i^2 +r_j^2} \right)^k  \label{eq:0b}
  \end{equation}
is a spherical Bessel-like function.
 In the independent particle approximation method, the potential function is given by 
 \begin{equation}
     \begin{split}
      V(\vec{r}_i, \vec{r}_j) &= -\frac{Z}{r_i} +  \sum_{j\neq i}^{n-1} \gamma_{l_i}\, \frac{1}{\mid \vec{r}_i-\vec{r}_j \mid}\\
      &=-\frac{Z}{r_i} + \sum_{j\neq i}^{n-1} \gamma_{l_i}\, \frac{4 \pi}{\sqrt{r_i^2+r_j^2}}\sum_{l,m} \tilde{j}_l(r_i,r_j)\, Y_l^{m*}(\hat{r_i})\,Y_l^{m}(\hat{r_j}),  \label{eq:pt1}
     \end{split}
    \end{equation}
for the $i^{\mathrm{th}}-$electron of the system. The coefficient $\gamma_{l_i}$ defines the ratio for partioning the correlation energy. Conventionally, factor $1/2$ which assumes equal sharing of the correlation energy between the interacting electrons is usually preferred. The interaction potential $ V(r_i, r_j)$ can be completely separated, by minimizing it with respect to the spatial co-ordinates.

  The spatially-dependent classical partitioning fraction 
 \begin{equation}
 \gamma_i (r_i,r_j) = \frac{r_i^2}{r_i^2 + r_j^2}, \label{eq:2}
 \end{equation}
 introduced in ref.\cite{Jobunga2017b}, is replaced with a corresponding symmetry-dependent partitioning fraction \cite{Jobunga2018, Jobunga2020} which depends, not on the radial coordinates, but on the local orbital angular momentum value ($l_i$) for the $i^{\mathrm{th}}$ state of the system. We employed an empirical process \cite{Jobunga2018} of obtaining the symmetry-dependent partitioning fractions used in this study.  The potentials developed in the previous paper \cite{Jobunga2017c} and related literature data aid in validating and calibrating the partitioning fractions. Indeed, the success of the equal sharing of the electron-electron interaction potential for spherically symmetric cases observed in ref. \cite{Jobunga2017c} already hinted to a possibility of existence of such a symmetry-dependent partitioning fraction. This hint was successfully used in coming up with a symmetry-dependent partitioning fraction for helium atom \cite{Jobunga2018, Jobunga2020}.

In our working, we established that the symmetry-dependent partitioning fraction is system specific and largely takes a general form which depends on the symmetry of the valence electrons. For valence electrons with an orbital angular momentum value $l_v$ and an orbital angular momentum $l_i$ for the $i^{\mathrm{th}}$ electron, the partitioning fraction $\gamma_{l_i}$ has been empirically determined \cite{Jobunga2020} to be
 \begin{equation}
 \displaystyle
 \gamma_{l_i} = \frac{1 + \delta_i}{2 + \delta_i+\delta_j }\\
  \label{eq:3a}
 \end{equation}
 where the parameters are given by 
 \begin{equation}
    \begin{split}
    \delta_i &= \left\{ \begin{matrix}
    0 & \mathrm{if }& l_i=0 \\ \sqrt[l_i]{l_i} & \mathrm{if }& l_i \neq 0 \end{matrix} \right.
    \end{split} \label{eq:3b}
    \end{equation} 
 and 
 \begin{equation}
 \delta_{j}= \frac{l_v}{2^{(2l_v-1)}} \label{eq:3c}
 \end{equation}
respectively. The non-local parameter $\delta_{j}$ depends on the valent subshell with $l_v$ being the orbital angular momentum quantum number of the valent electron for the atom in its groundstate configuration.

To simplify Eq. \eqref{eq:pt1} further, We utilise a mean-field approximation, developed in ref. \cite{Jobunga2020}, to evaluate the symmetry-dependent single-electron multipole potential 
\begin{equation}
V(\vec{r}_i) = \sum_{l=0}^{l_{\mathrm{max}}\rightarrow \infty} \sum_{m=-l}^{l} V_l^m(\vec{r}_i),\label{eq:4}
\end{equation}
for the two-electron system  where
  \begin{equation}
  \begin{split}
  V_{0}^0 (\vec{r}_i) &= -\frac{\left(Z-(n-1)\gamma_{l_i}\,B_0^0(Z)\,\sqrt[3]{\frac{Z}{(n-1)\gamma_{l_i}}}\right)}{r_i},\\
  V_l^m(\vec{r}_i) &= (n-1)\gamma_{l_i}\, B_l^m(Z)\,\sqrt[3]{\frac{Z}{(n-1)\gamma_{l_i}}}\,\frac{\exp\left[-\frac{lZr}{(l_i+1)(l_i-l+1)}\right]}{r_i^{l+1}}\, \delta_{l_i,l_j+l} \delta_{m_i,m_j+m}, \label{eq:5}
  \end{split}
  \end{equation}

 and the coefficient

 \begin{equation}
 B_l^m (Z)  = \frac{(2+2l_i-l)!}{(2l_i-2l+2)!}\; \left(\frac{2Z}{l_i-l+1} \right)^{(2l_i-2l+3)}\,\left[\frac{(l_i+1)(l_i-l+1)}{(2+2l_i-l)Z}\right]^{2l_i-l+3} A_l^m\langle \tilde{j}_l (r_i,r_j) \rangle.  \label{eq:6}
 \end{equation}
 The angular factors, 
 \begin{equation}
\begin{split}
A_l^m &= \int Y_{l_j}^{m_j}Y_{l}^{m*}Y_{l_i}^{m_i} \mathrm{d} \Omega_i \times \int Y_{l_i}^{m_i}Y_{l}^{m}Y_{l_j}^{m_j} \mathrm{d} \Omega_j   \\
&= (-1)^{m}\,\frac{(2l_i+1)(2l+1)(2l_j+1)}{4\pi}\left(\begin{matrix}l_i &l & l_j\\0&0&0 \end{matrix}\right)^2\, \left(\begin{matrix}l_i &l & l_j\\m_i&m&m_j \end{matrix}\right)^2, \label{eq:7}
\end{split}
\end{equation}
  in Eq. \eqref{eq:6} arise when the corresponding double integral for the correlated two-electron is solved. The expectation values 
  \begin{equation}
\langle \tilde{j}_l (r_i,r_j) \rangle = \left(\frac{1}{4 \pi \sqrt{2}}\right)^l \sum_{k=0,2,\cdots}^{\infty} \frac{(2l+2k-1)!!}{k!!(2l+k+1)!!}\left(\frac{1}{4 \pi \sqrt{2}}\right)^k, \label{eq:8}
\end{equation}
are approximated using a mean value, that is, the squareroot of its peak value per solid angle.
   Hydrogenic orbitals with angular momentum quantum number $l'$ and principal quantum number $l'+1$, where ${l'=l_i\; \mathrm{or}\; l_j}$, have been used in Eq. \eqref{eq:5} as the trial wavefunctions to solve the non-local exchange integral. The higher-order multipole potentials $V_{l}^m (\vec{r}_i)$ can then be added perturbatively to increase the accuracy of the single-electron potential. 
 
 The summation of all orders of the multipole potential in  Eqs. \eqref{eq:4} and \eqref{eq:5} simplify to a fully analytical single-electron potential function given by 
  \begin{equation}
   V(r_i) = -\frac{(Z-\sigma_{l_i})}{r_i}  \label{eq:9}
   \end{equation}
  where the electron screening parameter ${\sigma_{l_i}}$ is evaluated as
 \begin{equation}
 \sigma_{l_i} = (n-1)\gamma_{l_i}\,\sqrt[3]{\frac{Z}{\gamma_{l_i}}}\,\sum_{l=0}^{l_{\mathrm{max}}\rightarrow \infty} \sum_{m=-l}^{+l}   B_l^m (Z)\,\frac{\exp\left[-\frac{lZr}{(l_i+1)(l_i-l+1)}\right]}{r_i^l}\, \delta_{l_i,l_j+l} \delta_{m_i,m_j+m} . \label{eq:10}
 \end{equation}

 Using the suggested symmetry-dependent all-electron potential, the one electron Hamiltonian 
 \begin{equation}
 h_{\infty}(r_i)= \frac{p_i^2}{2} + V(r_i) \label{eq:11}
 \end{equation}
 
 without the finite nuclear mass correction is defined. With the finite nuclear mass correction, the Hamiltonian becomes
\begin{equation}
h(\vec{r}_i) = h_{\infty}(\vec{r}_i) -\frac{1}{M_{ion}} h_{\infty}(\vec{r}_i) \label{eq:11b}
\end{equation}
where ${1/M_{ion}}$ is the electron-atomic nuclear mass ratio.
 
 The eigenvalue $\epsilon_{\alpha_i}$ corresponding to state $\alpha_i$ for an $n$-electron atom can be generalised as
\begin{equation}
   \epsilon_{\alpha_i} = \frac{m}{n}\, \langle \phi_{\alpha_i}| h(\mathbf{r}_i)| \phi_{\alpha_i} \rangle \label{eq:12}
\end{equation} 
where ${m/n}$ refers to the proportion of non-vanishing integrals out of the possible permutations. For lithium, ${m/n=2/3}$ as was already shown in our previous paper \cite{Jobunga2017c}.  In principle, the integer $m$ can be determined from the spin-allowed groundstate configuration of the atom but constrained further by other symmetry considerations. We have provided a supplementary material \cite{Jobunga2021} showing how the various values of ${m/n}$ have been determined for up to ${n=7}$. For atoms with higher number of electrons, we have used an intuitive reasoning based on the arrangement of the systems in the periodic table to determine the value of ${m/n}$. A complete understanding of the symmetry relations for such large systems is, however, still necessary.   The ratio ${1:m-1}$ corresponds to the contribution of the direct and exchange integrals that contribute to the energy of the system respectively. This shows that apart from helium and alkalis where the exchange integral has an equal weight with the direct integral, in all the other atoms the exchange contribution is greater. Apparently, $m$ has a maximum value of $5$ for noble gases regardless of the number of electrons present in the atom.
 
For comparison purposes, we have also included results calculated using our previously derived lowest-order non-relativistic central potential \cite{Jobunga2017c}
\begin{equation}
 V(r_i) = -\frac{Z}{r_i} + (n-1)\,\frac{\left[\frac{Z\, f(r_i,r_j)}{2(n-1)}\right]^{3/5}}{r_i} \label{eq:13}
 \end{equation}
 where the expectation value  
 \begin{equation}
 \langle f(r_i, r_j)^{\frac{3}{5}} \rangle \approx 1 - \left[ \frac{27}{25} + \frac{3}{5}Zr_i -\frac{6}{125\,Zr_i}\right]\, \exp (-2Zr_i) \label{eq:14}
 \end{equation}
 is approximately optimized by evaluating the integral using a trial function for hydrogenic system in the $1s$ state.
 
\section{Results and Discussions}

We have developed a symmetry-dependent all-electron potential for an $n$-electron system defined by Eqs. (\ref{eq:4}) and (\ref{eq:5}). The potentials are used to calculate the groundstate ionization potentials for $n$-electron atoms as shown in tables \ref{tab1} and \ref{tab1b} with ${2\leq n \leq 103}$. Our results are compared with the results of the central potential given by Eq. \eqref{eq:13}, DFT calculations \cite{Kraisler2010}, and experimental reference data \cite{NIST}. In generating our results, a B-spline radial box of $600$ B-splines, maximum radius ${r_{\mathrm{max}}=200}$, order $k=10$, and a non-linear knot sequence is used. Our results can also be evaluated analytically, if the suggested all-electron potential is restricted to the spherical terms only, by determining the nuclear screening constants and the effective nuclear charge.

\begin{table}[H]
    \centering
    \begin{tabular}{ccccccccccccc}
    \hline
 $n$ & Atom&State& $m/n$ &$V_{\mathrm{cen}}$&$V_{0,0}^0$ &$V_0$& $V_{h}$&$V_{\mathrm{fnm}}$ &LSDA&GGA &Exp.(eV)&$\Delta \varepsilon$\\
   \hline
   \hline
       $2$& He&1s&2/2   &35.21 & 24.77 &24.60 & 24.60 & 24.59 &24.29&24.46& 24.58& +0.01 \\                                                     
       $3$& Li&2s&2/3   &4.97 & 5.50   &5.48 &5.48 & 5.48 &5.47&5.58& 5.39& +0.09 \\                          
       $4$& Be&2s&3/4   &8.91  & 9.40  &9.37 &9.37 & 9.36 &9.02&8.99& 9.32& +0.04 \\  
       $5$& B &2p&3/5   &8.08 & 8.43   &8.39 &8.35 & 8.35 &8.57&8.48& 8.29 & +0.06\\ 
       $6$& C &2p&4/6   &12.29 &12.42  &12.36 &12.31 & 12.30 &11.76&11.67& 11.26& +1.04\\                                  
       $7$& N &2p&4/7   &13.83 &13.65  &13.58 &13.52 & 13.51 &14.99&14.91& 14.53& -1.02\\                                                     
       $8$& O &2p&4/8   &15.37 &14.90  &14.82 &14.76 & 14.75 &13.89&13.67& 13.61& +1.14\\                          
       $9$& F &2p&4/9   &16.93 &16.16  &16.07 &16.01 & 16.00 &18.05&17.84& 17.42& -1.42\\  
       $10$&Ne&2p& 5/10 &23.12 &21.80  &21.68 &21.59 & 21.58 &22.17&21.98& 21.56& +0.02\\ 
       $11$&Na&3s& 2/11 &5.30 & 5.55   &5.53 &5.53 &5.52 &5.36&5.35& 5.13 & +0.39\\                                  
       $12$&Mg&3s& 3/12 &8.59 & 8.95   &8.90 &8.90 &8.90 &7.60&7.71&7.64& +1.26 \\
       $13$& Al&3p&3/13 &7.72 &7.10   &7.06  &7.04 & 7.03 &5.99&5.96& 5.98& +1.05 \\                                                     
       $14$& Si&3p&4/14 &10.99 &10.05 &9.99  &9.96 & 9.95 &8.27&8.27& 8.15& +1.80 \\                          
       $15$& P&3p&4/15  &11.69 &10.63 &10.56  &10.53 & 10.52 &10.53&10.55& 10.48& +0.04 \\  
       $16$& S &3p&4/16 &12.39 &11.20 &11.14 &11.10 & 11.09 &10.53&10.23& 10.36 & +0.73\\ 
       $17$& Cl &3p&4/17&13.09 &11.79 &11.71  &11.68 & 11.67 &13.24&13.05& 12.96& -1.29\\                                  
       $18$& Ar &3p&5/18 &17.24 &15.46 &15.36  &15.32 & 15.30 &15.93&15.80& 15.75& -0.45\\                                                     
       $19$& K &4s&2/19 &4.65  &4.97   &4.94 &4.94 & 4.94 &4.52&4.44& 4.34 &+0.60 \\                          
       $20$& Ca &4s&3/20&7.31  &7.80  &7.77  &7.77 & 7.75 &6.20&6.06& 6.11& +1.64 \\  
       $21$&Sc&3d& 3/21 &11.86 &6.79  &6.73  &6.71 & 6.70 &6.54&6.37& 6.56 & +0.14\\ 
       $22$&Ti&3d& 3/22  &12.39 &7.05 &6.99 &6.97  & 6.96 &6.71&6.54& 6.82& +0.14 \\                                  
       $23$&V &3d&3/23   &12.91 &7.32 &7.25  &7.24 & 7.22 &7.14&6.96&6.74& +0.48\\                          
       $24$&Cr &3d&3/24   &13.43 &7.58 &7.51  &7.50 & 7.48 &7.45&7.25&6.76 & +0.72\\  
       $25$&Mn&3d& 3/25 &13.96 &7.85  &7.78  &7.76 & 7.75 &7.48&7.15&7.43 & +0.32\\ 
       $26$&Fe&3d& 3/26 &14.48 &8.11  &8.04  &8.02 & 8.01 &8.20&7.84&7.90 & +0.11\\                                  
       $27$&Co&3d& 3/27 &15.01 &8.38  &8.30  &8.28 & 8.27 &8.10&7.89&7.88 & +0.39\\
       $28$&Ni&3d&3/28 &15.53 & 8.64 &8.57  &8.55 & 8.53 &8.23&8.01&7.63 & +0.90\\                                                     
       $29$&Cu&3d&3/29 &16.06 &8.91  &8.83  &8.81 & 8.79 &8.38&8.14&7.72 & +1.07\\                          
       $30$&Zn&3d&3/30  &16.58 &9.17 &9.09  & 9.07 & 9.05 &9.68&9.37&9.39& -0.34 \\  
       $31$&Ga&4p&3/31 &9.66 &8.41  &8.36  &8.34 & 8.32 &6.05&5.90&5.99& +2.33 \\ 
       $32$&Ge&4p&4/32&13.27 &11.55  &11.47  &11.45 & 11.42 &8.06&7.97&7.89&+3.53 \\                                  
       $33$&As&4p&4/33 &13.67 &11.88 &11.80  &11.78 & 11.75 &9.97&9.93&9.78& +1.97\\                                                     
       $34$&Se&4p&4/34 &14.06  &12.21 &12.13 &12.10 & 12.07 &9.92&9.54&9.75 & +2.32\\                          
       $35$&Br&4p&4/35&14.46  &12.54  &12.45 &12.43 & 12.40 &12.12&11.87&11.81& +0.59 \\  
       $36$&Kr&4p& 5/36 &18.56 &16.08 &15.98 &15.94 & 15.90 &14.26&14.09&13.99& +1.91\\                           
       $37$&Rb&5s&2/37&5.42  &5.85  &5.82  &5.82 & 5.81 &4.32&4.21&4.17 & +1.64\\  
       $38$&Sr&5s& 3/38 &8.35 &9.01  &8.96  &8.96 & 8.94 &5.77&5.61&5.69& +3.25 \\
       $39$& Y &4d&3/39   &11.98 &6.51 &6.45  &6.43 & 6.42 &6.24&6.02&6.21 & +0.21\\                                                     
       $40$& Zr &4d&3/40  &12.27 & 6.66&6.59  &6.58 & 6.56 &6.68&6.48&6.63& -0.07 \\                          
       $41$& Nb &4d&3/41  &12.57 &6.81 &6.74  &6.73 & 6.71 &7.03&6.80&6.75 & -0.04\\  
       $42$& Mo &4d&3/42  &12.86 & 6.96 &6.89 &6.88 & 6.86 &7.29&7.04&7.09& -0.23 \\ 
       $43$& Tc &4d&3/43  &13.16 & 7.11 &7.04 &7.03 & 7.01 &7.45&7.21&7.11& -0.10\\                                  
       $44$& Ru &4d&3/44  &13.46 & 7.26 &7.19 &7.17 & 7.15 &7.54&7.29&7.36& -0.21 \\                                                     
       $45$& Rh &4d&3/45  &13.75 & 7.41 &7.34 &7.32 & 7.30 &8.29&7.97&7.45& -0.15\\                          
       $46$& Pd &4d&3/46  &14.05 & 7.56 &7.49 &7.47 & 7.45 &9.37&9.08&8.33& -0.88\\  
       $47$& Ag &4d&3/47  &14.34 & 7.71 &7.63 &7.62 & 7.60 &7.66&7.36&7.57& +0.03\\ 
       $48$& Cd &4d&3/48  &14.64 &7.86 &7.78 &7.77 & 7.74 &8.85&8.50&8.99 & -1.25\\                                  
       $49$& In &5p&3/49  &9.58  &8.23 &8.17  &8.16 & 8.13 &5.77&5.58&5.78&  +2.35\\
       $50$& Sn &5p&4/50  &13.03 & 11.18&11.11 &11.09 & 11.05 &7.48&7.36&7.34& +3.71 \\                                                     
       $51$& Sb &5p&4/51  &13.28 & 11.39&11.32 &11.30 & 11.26 &9.08&9.01&8.60& +2.66\\                                                     
       $52$& Te &5p&4/52  &13.54 & 11.60&11.53 &11.51 & 11.47 &9.06&8.69&9.00& +2.47\\                          
       $53$& I &5p&4/53   &13.79 & 11.82&11.74 &11.72 & 11.67 &10.85&10.59&10.45& +1.22\\  
       $54$& Xe &5p&5/54  &17.55 & 15.03&14.93 &14.91 & 14.85 &12.56&12.38&12.12& +2.73 \\ 
    \hline
    \end{tabular}
    \caption{Some numerically calculated non-relativistic ionisation potentials (in $eV$) for $2\le n \le 54$-electron atoms versus the reference values from DFT (LSDA and GGA) calculations \cite{Kraisler2010} and experiment \cite{NIST}.  The Present $V_{cent}$ results are evaluated using Eq. \eqref{eq:13} and $V_{0,0}^0$ are evaluated using the lowest-order symmetry-dependent non-relativistic potential given by Eq. \eqref{eq:5}  with ${j_0(r_i,r_j) \approx 1}$. The present $V_0$ and $V_h$ are the results evaluated using the symmetry-dependent all-electron multipole potential given by Eqs. \eqref{eq:4} and \eqref{eq:5} by fixing ${l_{\mathrm{max}}=0}$ and ${l_{\mathrm{max}}=4}$ respectively. The present $V_{\mathrm{fnm}}$ results are obtained by including the finite nuclear mass scaling corrections, using Eq. \eqref{eq:11b}, on the multipole $V_h$ data. The results presented are truncated to $2$ d.p. The deviations ${\Delta \varepsilon}$ are also determined by subtracting the experimental results from our $V_{\mathrm{fnm}}$ results.}
    \label{tab1}
  \end{table}
\begin{table}[H]
    \centering
    \begin{tabular}{cccccccccccccc}
    \hline
 $n$ & Atom&State& $m/n$ &$V_{\mathrm{cen}}$&$V_{0,0}^0$ &$V_0$& $V_{h}$&$V_{\mathrm{fnm}}$ &LSDA&GGA& Exp.(eV)&$\Delta \epsilon$ \\
   \hline
   \hline
       $55$& Cs &6s&2/55  &5.42  & 5.93 &5.89 & 5.89 & 5.87 &3.94&3.83&3.89& +1.98\\                                  
       $56$& Ba &6s&3/56  &8.28  & 9.05 &9.00 & 9.00 & 8.96 &5.16&4.97&5.21& +3.75\\                                                     
       $57$& La &4f&3/57  &17.29 & 7.54 &7.46 &7.45 & 7.42 &5.31&4.98&5.57& +1.85\\                          
       $58$& Ce &4f&3/58  &17.59 & 7.66 &7.58 &7.57 & 7.54 &5.12&4.96&5.53& +2.01 \\  
       $59$& Pr &4f&3/59  &17.88 & 7.78 &7.70 &7.69 & 7.66 &4.99&4.82&5.47& +2.19 \\ 
       $60$& Nd &4f&3/60  &18.18 & 7.91 &7.82 &7.81 & 7.78 &4.88&4.70&5.52& +2.26 \\                                  
       $61$& Pm &4f&3/61  &18.47 & 8.03 &7.94 &7.93 & 7.90 &4.78&4.60&5.58& +2.32\\                          
       $62$& Sm &4f&3/62  &18.77 & 8.15 &8.06 &8.05 & 8.02 &4.70&4.53&5.64& +2.38\\  
       $63$& Eu &4f&3/63  &19.06 & 8.27 &8.18 &8.17 & 8.13 &5.51&5.33&5.67& +2.46\\ 
       $64$& Gd &4f&3/64  &19.36 & 8.39 &8.30 &8.29 & 8.25 &5.60&5.42&6.14& +2.11\\                                  
       $65$& Tb &4f&3/65  &19.65 & 8.51 &8.42 &8.41 & 8.37 &5.49&5.35&5.86 & +2.51\\
       $66$& Dy &4f&3/66  &19.95 & 8.64 &8.54 &8.53 & 8.49 &5.40&5.26&5.93 & +2.56\\                                                     
       $67$& Ho &4f&3/67  &20.24 & 8.76 &8.66 &8.65 & 8.61 &5.32&5.17&6.02& +2.59 \\                          
       $68$& Er &4f&3/68  &20.54 & 8.88 &8.78 &8.77 & 8.73 &5.26&5.10&6.10 & +2.63\\  
       $69$& Tm &4f&3/69  &20.84 & 9.00 &8.90 &8.89 & 8.85 &5.20&5.07&6.18& +2.67\\ 
       $70$& Yb &4f&3/70  &21.13 & 9.12 &9.02 &9.01 & 8.97 &6.00&5.82&6.25& +2.72\\                                  
       $71$& Lu &5d&3/71  &13.71 & 7.24 &7.17 &7.16 & 7.12 &6.39&6.18&5.42& +1.70\\                                                     
       $72$& Hf &5d&3/72  &13.90 &7.33  &7.26 &7.25 & 7.22 &6.76&6.59&6.82 & +0.40\\                          
       $73$& Ta &5d&3/73  &14.09 &7.43  &7.36 &7.35 & 7.31 &7.23&7.01&7.54 & -0.23\\  
       $74$& W &5d&3/74   &14.28 &7.53  &7.45 &7.44 & 7.40 &7.49&7.25&7.86 & -0.46\\                           
       $75$& Re &5d&3/75  &14.47 &7.62  &7.55 &7.54 & 7.50 &7.63&7.15&7.83& -0.33\\  
       $76$& Os &5d&3/76  &14.66 &7.72  &7.64 &7.63 & 7.59 &7.69&7.45&8.43 & -0.84\\
       $77$& Ir &5d&3/77 &14.84 &7.81  &7.74 &7.73 & 7.69 &8.57&8.24&8.96 & -1.27\\                                                     
       $78$& Pt &5d&3/78 &15.03 &7.91  &7.83 &7.82 & 7.78 &9.67&9.38&8.95& -1.17 \\                          
       $79$& Au &5d&3/79 &15.22 &8.01  &7.93 &7.92 & 7.87 &7.75&7.45&9.22& -1.35 \\  
       $80$& Hg &5d&3/80 &15.41 &8.10  &8.03 &8.01 & 7.97 &8.82&8.46&10.43 & -2.46 \\ 
       $81$& Tl &6p&3/81 &10.86 &9.23  &9.16 &9.15 & 9.10 &5.67&5.46&6.10  & +3.00\\                                  
       $82$& Pb &6p&4/82 &14.66 &12.45 &12.37 &12.35 & 12.28 &7.25&7.12&7.41  & +4.87\\                                                     
       $83$& Bi &6p&4/83 &14.83 &12.60 &12.51 &12.49 & 12.42 &8.71&8.63&7.28 & +5.14\\                          
       $84$& Po &6p&4/84 &15.01 &12.74 &12.66 &12.64 & 12.57 &8.69&8.32&8.41  & +4.16\\  
       $85$& At &6p&4/85 &15.18 &12.89 &12.80 &12.79 & 12.71 &10.29&10.02&9.31 & +3.40\\ 
       $86$& Rn &6p&5/86 &19.20 &16.30 &16.19 &16.16 & 16.07 &11.81&11.63&10.74  & +5.33\\                                  
       $87$& Fr &7s&2/87 &6.15  &6.78  &6.75 &6.75 & 6.71 &&&4.07 & +2.64\\
       $88$& Ra &7s&3/88 &9.33  &10.29 &10.24 &10.24 & 10.18 &&&5.27& +4.91\\                                                     
       $89$& Ac &5f&3/89 &17.11 &7.32 &7.24 &7.23  & 7.18 &&&5.38& +1.80\\                          
       $90$& Th &5f&3/90 &17.30 &7.40  &7.32 & 7.30 & 7.26 &&&6.30& +0.96\\  
       $91$& Pa &5f&3/91 &17.49 &7.47 &7.39  & 7.38 & 7.34 &&&5.89& +1.45\\ 
       $92$& U  &5f&3/92 &17.68 &7.55 &7.47  & 7.46 & 7.41 &&&6.19& +1.22\\                                  
       $93$& Np &5f&3/93 &17.87 &7.63 &7.55  & 7.54 & 7.49 &&&6.26& +1.23\\                                                     
       $94$& Pu &5f&3/94 &18.06 &7.71  &7.62 & 7.61 & 7.56 &&&6.02& +1.54\\                          
       $95$& Am &5f&3/95 &18.25 &7.79 &7.70  & 7.69 & 7.64 &&&5.97& +1.67\\  
       $96$& Cm &5f&3/96 &18.44 &7.86 &7.78  & 7.77 & 7.72 &&&5.99& +1.73\\ 
       $97$& Bk &5f&3/97 &18.62 &7.94 &7.86  & 7.84 & 7.79 &&&6.19& +1.60 \\                                  
       $98$& Cf &5f&3/98 &18.81 &8.02 &7.93  & 7.92 & 7.87 &&&6.28& +1.59\\                          
       $99$& Es &5f&3/99 &19.00 &8.10 &8.01 & 8.00 & 7.94 &&&6.36& +1.58\\  
       $100$& Fm &5f& 3/100 &19.19 &8.18&8.09 &8.07 & 8.02 &&&6.50& +1.52\\ 
       $101$& Md &5f& 3/101 &19.38 &8.25&8.16 &8.15 & 8.10 &&&6.58& +1.52\\                                  
       $102$& No &5f& 3/102 &19.57 &8.33&8.24 &8.23 & 8.17 &&&6.62& +1.55\\
       $103$& Lr &6d& 3/103 &13.72 &7.16&7.09 &7.08 & 7.03 &&&4.96&+2.07\\
    \hline
    \end{tabular}
    \caption{Same as table \ref{tab1} but for $55 \le n \le 103$-electron atoms. }
    \label{tab1b}
  \end{table}

Tables \ref{tab1} and \ref{tab1b} show some numerically calculated non-relativistic ionization energies for $n$-electron atoms using the present central lowet-order potential $V_{\mathrm{cen}}$ given by Eq. \eqref{eq:13} and the symmetry-dependent all-electron potentials in Eqs. \eqref{eq:4} and \eqref{eq:5}, evaluated by fixing ${j_0(r_i,r_j) \approx 1}$ for the lowest-order $V_{0,0}^0$ term, and ${l_{\mathrm{max}}=0}$ and ${l_{\mathrm{max}}=4}$  for the higher order $V_0$ and $V_h$ terms respectively. We have also included the finite nuclear mass corrections in our higher multipole order ${V_{\mathrm{fnm}}}$ results. Our results are compared with the reference theoretical \cite{Kraisler2010} and experimental values \cite{NIST}. The tables also contain the groundstate valent orbital, the  energy deviation ($\Delta \epsilon$) from the experimental values as well as the single-electron Hamiltonian scaling fractions $(m/n)$ for each of the atoms, already introduced in Eq. \eqref{eq:12}, that have been determined and used in the present study. These scaling fractions yield information on the relative importance of the exchange integrals, arising from the permutation symmetry, in the evaluation of the energy eigenvalues.   
  
The ionisation energies have been obtained using Koopman's theorem \cite{KOOPMANS1934} except for the groundstate of helium atom, which is evaluated as the difference between the total energy of helium atom and the residual ion both in their groundstates.
The groundstate ionisation energies calculated using the present all-electron potential incorporating the multipole terms, without ($V_{h}$) and with the finite nuclear mass correction (${V_{\mathrm{fnm}}}$), as given by Eqs.\eqref{eq:4}, \eqref{eq:5}, \eqref{eq:11} and \eqref{eq:11b} are in fair agreement with the reference values.
We have also included the results generated using the lowest-order central ($V_{\mathrm{cen}}$) and symmetry-dependent ($V_{0,0}^0$) non-relativistic potential as well as a higher-order multipole spherical ($V_0$) potential. With these additional potentials, the effects of the dominant lowest-order interaction, the central potential, the symmetry-dependent potential, the higher-order spherical and non-spherical interactions, and of the finite nuclear mass corrections can be investigated.

The central potential ($V_{\mathrm{cen}}$) given by Eq. \eqref{eq:13} and the symmetry-dependent potential ($V_{0,0}^0$) interaction terms  are both lowest-order non-relativistic interaction terms of the multipole interaction potential, with the difference between them being the nature of the partioning fraction (${\gamma_{l_i}}$) used for each case. In one case, a radial partitioning fraction is used while in another case, a symmetry-dependent partitioning fraction is used in the treatment of the electron-electron interaction terms. The results between the two corresponding potentials compare well qualitatively but the quantitative difference increase with the atomic nuclear charge $(Z)$.  The discrepancy between the results generated by the lowest multipole-order based potentials can be attributed to the approximation involved in the central potential given by Eq. (\ref{eq:14}). Since the calculations involving the symmetry-dependent partitioning fraction can be considered to be exact, without any approximation used, their values can be considered to be reliable within the non-relativistic regime subject to the validity of the use of both the partitioning and scaling fractions. 
In deed if the higher-order  interactions and finite nuclear mass corrections are included into the symmetry-dependent single-electron Hamiltonian, there is some slight improvement in the agreement of our results with the experimental values for all the multi-electron atoms. The trend of the ionisation energies in the periodic table is correctly predicted by our all-electron potential. The quantitative agreement with experimental values is better for few-electron atoms but decrease with higher $Z$ values. 
Some major discrepancies are, however, observed between our evaluated results and the experimental results for some of the multi-electron atoms across the $Z$ spectrum. For example, our potential is unable to explain why the ionisation energy of oxygen atom is lower than that of nitrogen atom. The source of such discrepancies is not quite clear but some can be attributed to the dominant role played by the relativistic effects to some extent, especially for high $Z$ atoms. It is of desirable interest to include the effects arising from the relativistic and other higher-order interactions in our future investigation.

\begin{figure}[H]
  \center
    \includegraphics*[width= 0.75\linewidth]{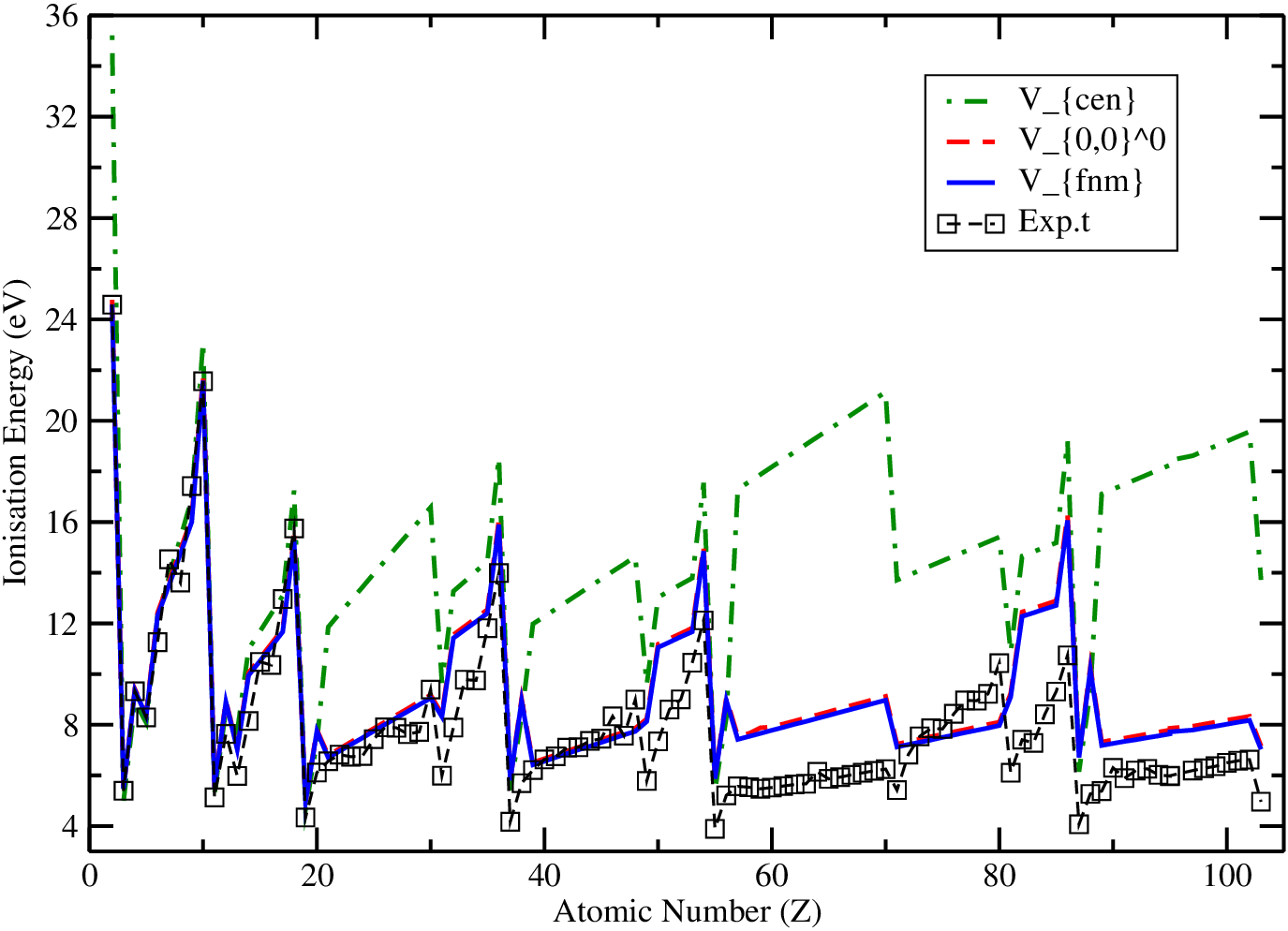}
    
    \caption{(Color online) Non-relativistic atomic ionization energies for multi-electron atoms evaluated using various potentials in comparison with experimental results \cite{NIST}. }
  \label{fig:1}
\end{figure}

\begin{figure}[H]
  \center
    \includegraphics*[width= 0.75\linewidth]{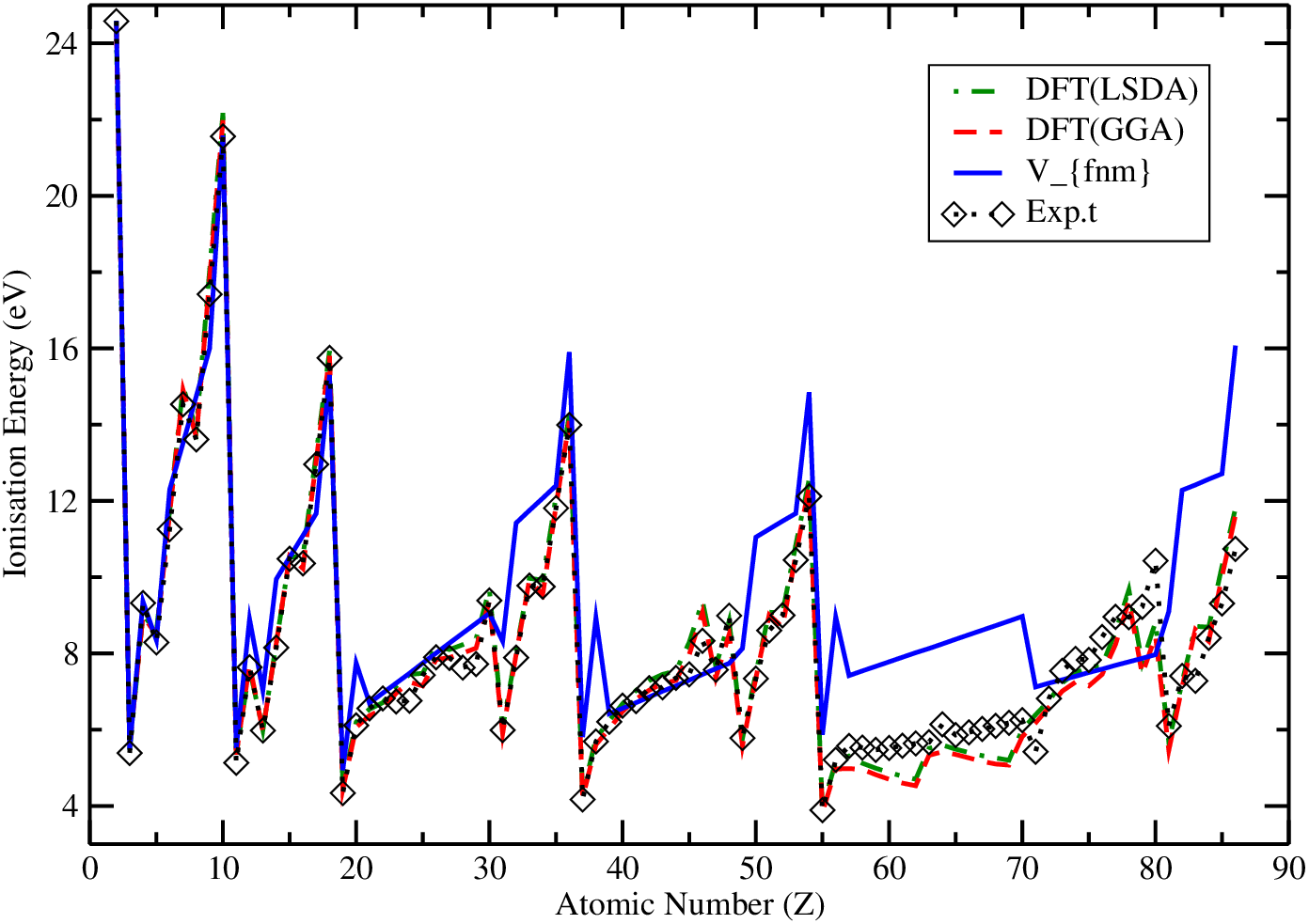}
    
    \caption{(Color online) Non-relativistic atomic ionisation energies for multi-electron atoms evaluated using our symmetry-dependent higher multipole-order potential, with finite nuclear mass corrections,  in comparison with DFT (LSDA and GGA) \cite{Kraisler2010} and experimental results \cite{NIST}. }
  \label{fig:2}
\end{figure}

Figures \ref{fig:1} and \ref{fig:2} graphically represents the results contained in tables \ref{tab1} and \ref{tab1b}. In Fig. \ref{fig:1}, our present results are compared with the experimental results derived from literature. It is evident from the graphs that all sets of results are in reasonable agreement up to ${Z=20}$. Beyond this point, the central potential results break down. This is because central potentials average results around the ${l=0}$ values, for equivalent principal quantum numbers, leading to bigger deviations for higher $l$ values. The nuclear charge $Z$ amplifies such deviations. The symmetry-dependent potential, on the other hand, follows a pattern which is unique for every $l$ value and for every atom. The symmetry-dependent results yield a reasonable prediction for the experimental values, but significant deviation from experiment persists even for this potential. It can be noticed that the influence of higher multipole corrections, beyond the lowest-order non-relativistic potential, is small.  

In Fig. \ref{fig:2}, our calculated results incorporating higher-order multipole potentials and finite nuclear mass corrections are compared with DFT calculations with LSDA and GGA exchange correlation functions \cite{Kraisler2010} and with the experimental \cite{NIST} results. Evidently, both sets of the DFT spin polarised results compare well with experimental results, but our calculated results are also in reasonable agreement, and even better for some many-electron atoms like helium, lithium, beryllium, boron, neon, phosphorus, and a few others. Our method is, nevertheless, not directly comparable with the DFT making it a challenging task to explain the disparity between them. 

\begin{figure}[H]
  \center
    \includegraphics*[width= 0.75\linewidth]{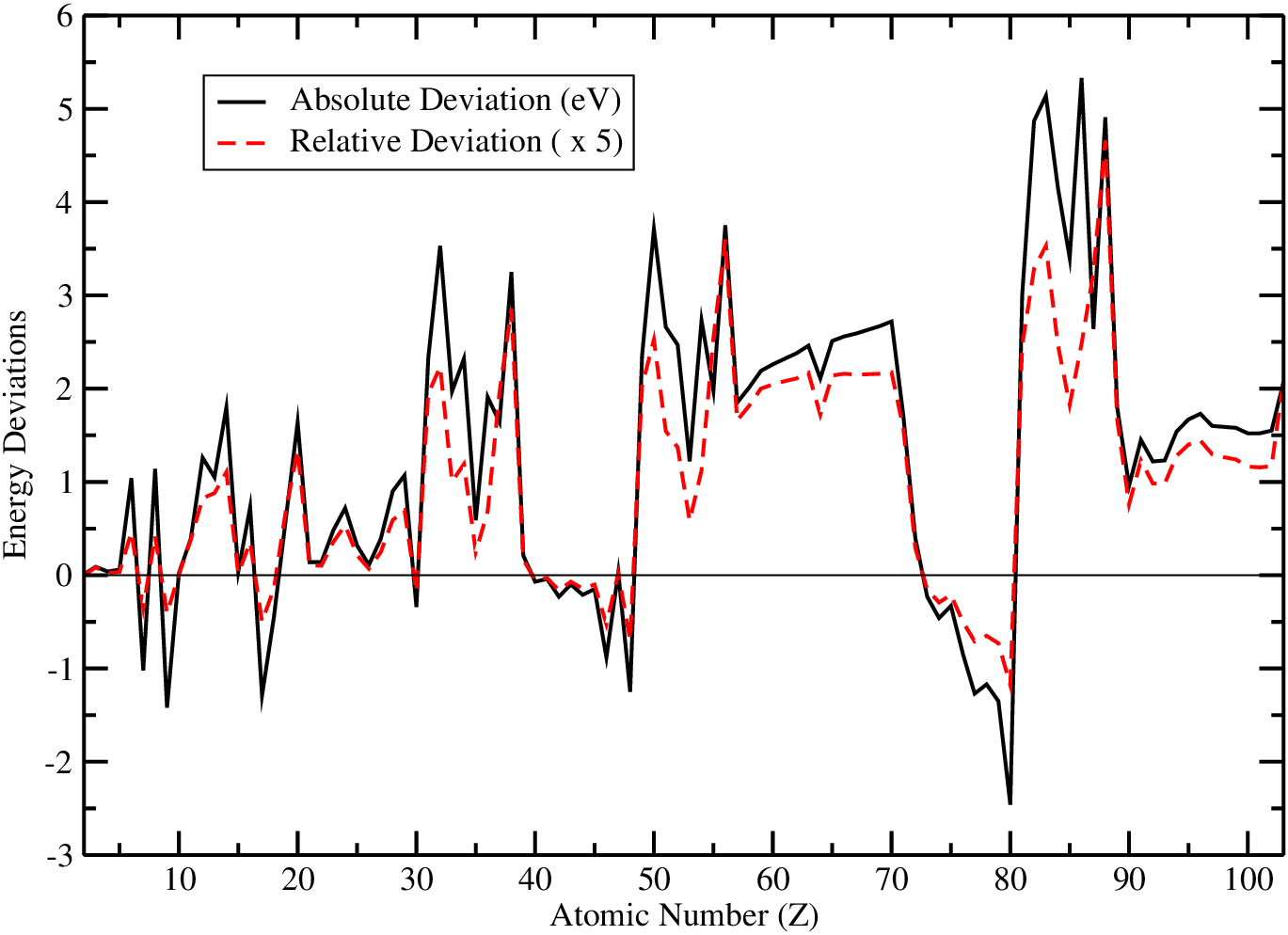}
    
    \caption{(Color online) Ionisation energies deviations, that is the absolute (${\mathrm{Abs. Dev.}= V_{\mathrm{fnm}}-\mathrm{Exp}}$ in eV) and relative deviations, from the experimental results  for the multi-electron atoms as a function of atomic number ($Z$). The relative deviations, multiplied by a factor of $5$, are obtained by dividing the absolute deviation with the experimental ionisation energies. Our higher-order multipole results with finite nuclear mass corrections have been considered to be more realistic for the calculations  of deviations.}
  \label{fig:3}
\end{figure}

Figure \ref{fig:3} shows the absolute and relative energy deviations from the experimental values. The absolute energy deviation is obtained by subtracting the experimental results from our higher-order multipole potential (${V_{\mathrm{fnm}}}$) results with finite nuclear mass corrections. The relative energy deviations, on the other hand, are obtained by dividing the absolute deviations with the experimental values. Our relative deviations have been multiplied by a factor of $5$ to fit on the same scale with the absolute deviation.
Positive deviation implies that our binding energies for the respective orbitals are lower than the experimental values while negative deviation implies that our binding energies are higher. From the figure, it can be observed that the deviations show a high degree of correlation between them. It is also apparent that both deviations increase with atomic number ($Z$) to some extent. In general, the disparities between our present results, the experimental results, and both sets of DFT results point to the need for further research incorporating relativistic and other higher-order corrections in order to clarify the uncertainties existing in both theory and experiment.

\section{Conclusion}
We have generalised the symmetry-dependent all-electron potential  for $n$-electron atoms. With this potential, all atoms are treated in the same way regardless of whether they are open- or closed-shell using only their system specific information. The scaling fractions ($m/n$) used have been derived for up to $7$-electron systems and extended empirically and using symmetry arguments for the rest of the $n$-electron systems. In this study, the permanent (symmetric form of Slater determinants) expansion of single electron spin-orbitals is used to express the wavefunction. This may be contrary to the fundamental postulates of quantum mechanics but, in a way, similar to the spatial expansion of the singlet states of helium atom using the Slater determinants. Our motivation for using such an expansion of the wavefunction was necessitated by the separability of the interaction potential leading to a form of a Hamiltonian for non-interacting fermions. The performance of the potential is tested against our previously developed central potential and DFT calculations in reproducing benchmark experimental results. The contributions of the lowest-order and higher-order non-relativistic interaction potentials in the various ionisation energies are evaluated. 
The potential yields reliable groundstate ionisation energies relative to literature data. The major advantage of the derived potential is that it leads to a completely separable Hamiltonian for the many-electron atoms. This eliminates the need for self consistent field iterations usually employed in other commonly used theoretical methods like the Hartree-Fock, configuration interaction (CI), and DFT. It hinges further on the premise that any two interacting electrons can exchange their relative position coordinates without exchanging their spin degrees of freedom. The suggested potential has a possibility of further improvement by incorporating the relativistic and other higher order interactions.

\section{Acknowledgement}
The author is grateful to the German Center for Migration and Development (CIM) for partial funding of the project under grant number $\mathrm{RE}50095158$.

\bibliographystyle{apsrev}
\bibliography{/home/eric/Inworks/Literature}

\begin{thebibliography}{23}
\expandafter\ifx\csname natexlab\endcsname\relax\def\natexlab#1{#1}\fi
\expandafter\ifx\csname bibnamefont\endcsname\relax
  \def\bibnamefont#1{#1}\fi
\expandafter\ifx\csname bibfnamefont\endcsname\relax
  \def\bibfnamefont#1{#1}\fi
\expandafter\ifx\csname citenamefont\endcsname\relax
  \def\citenamefont#1{#1}\fi
\expandafter\ifx\csname url\endcsname\relax
  \def\url#1{\texttt{#1}}\fi
\expandafter\ifx\csname urlprefix\endcsname\relax\def\urlprefix{URL }\fi
\providecommand{\bibinfo}[2]{#2}
\providecommand{\eprint}[2][]{\url{#2}}

\bibitem[{\citenamefont{Hugenholtz}(1965)}]{Hugenholtz1965}
\bibinfo{author}{\bibfnamefont{N.~M.} \bibnamefont{Hugenholtz}},
  \bibinfo{journal}{Reports on Progress in Physics}
  \textbf{\bibinfo{volume}{28}}, \bibinfo{pages}{201} (\bibinfo{year}{1965}).

\bibitem[{\citenamefont{Verdebout et~al.}(2013)\citenamefont{Verdebout, Rynkun,
  Jönsson, Gaigalas, Fischer, and Godefroid}}]{Verdebout2013}
\bibinfo{author}{\bibfnamefont{S.}~\bibnamefont{Verdebout}},
  \bibinfo{author}{\bibfnamefont{P.}~\bibnamefont{Rynkun}},
  \bibinfo{author}{\bibfnamefont{P.}~\bibnamefont{Jönsson}},
  \bibinfo{author}{\bibfnamefont{G.}~\bibnamefont{Gaigalas}},
  \bibinfo{author}{\bibfnamefont{C.~F.} \bibnamefont{Fischer}},
  \bibnamefont{and}
  \bibinfo{author}{\bibfnamefont{M.}~\bibnamefont{Godefroid}},
  \bibinfo{journal}{Journal of Physics B: Atomic, Molecular and Optical
  Physics} \textbf{\bibinfo{volume}{46}}, \bibinfo{pages}{085003}
  (\bibinfo{year}{2013}).

\bibitem[{\citenamefont{Tobocman}(1957)}]{Tobocman1957}
\bibinfo{author}{\bibfnamefont{W.}~\bibnamefont{Tobocman}},
  \bibinfo{journal}{Phys. Rev.} \textbf{\bibinfo{volume}{107}},
  \bibinfo{pages}{203} (\bibinfo{year}{1957}).

\bibitem[{\citenamefont{Cremer}(2013)}]{Cremer2013}
\bibinfo{author}{\bibfnamefont{D.}~\bibnamefont{Cremer}},
  \bibinfo{journal}{Wiley Interdisciplinary Reviews: Computational Molecular
  Science} \textbf{\bibinfo{volume}{3}}, \bibinfo{pages}{482}
  (\bibinfo{year}{2013}), ISSN \bibinfo{issn}{1759-0884}.

\bibitem[{\citenamefont{Kohn and Sham}(1965)}]{Kohn1965}
\bibinfo{author}{\bibfnamefont{W.}~\bibnamefont{Kohn}} \bibnamefont{and}
  \bibinfo{author}{\bibfnamefont{L.~J.} \bibnamefont{Sham}},
  \bibinfo{journal}{Phys. Rev.} \textbf{\bibinfo{volume}{140}},
  \bibinfo{pages}{A1133} (\bibinfo{year}{1965}).

\bibitem[{\citenamefont{Cramer}(2002)}]{Cramer2002}
\bibinfo{author}{\bibfnamefont{C.~J.} \bibnamefont{Cramer}},
  \emph{\bibinfo{title}{Essentials of Computational Chemistry}}
  (\bibinfo{publisher}{John Wiley and Sons, Inc},
  \bibinfo{address}{Chichester}, \bibinfo{year}{2002}).

\bibitem[{\citenamefont{Hylleraas}(1929)}]{Hylleraas1929}
\bibinfo{author}{\bibfnamefont{E.~A.} \bibnamefont{Hylleraas}},
  \bibinfo{journal}{Zeitschrift f{\"u}r Physik} \textbf{\bibinfo{volume}{54}},
  \bibinfo{pages}{347} (\bibinfo{year}{1929}).

\bibitem[{\citenamefont{Cohen}(1984)}]{Cohen1984}
\bibinfo{author}{\bibfnamefont{M.~L.} \bibnamefont{Cohen}},
  \bibinfo{journal}{Ann. Rev. Mater. Sci} \textbf{\bibinfo{volume}{14}},
  \bibinfo{pages}{119} (\bibinfo{year}{1984}).

\bibitem[{\citenamefont{Hellmann}(1935)}]{Hellmann1935}
\bibinfo{author}{\bibfnamefont{J.~J.} \bibnamefont{Hellmann}},
  \bibinfo{journal}{J. Chem Phys} \textbf{\bibinfo{volume}{3}},
  \bibinfo{pages}{61} (\bibinfo{year}{1935}).

\bibitem[{\citenamefont{Callaway and Laghos}(1969)}]{Callaway1969}
\bibinfo{author}{\bibfnamefont{J.}~\bibnamefont{Callaway}} \bibnamefont{and}
  \bibinfo{author}{\bibfnamefont{P.~S.} \bibnamefont{Laghos}},
  \bibinfo{journal}{Physical Review} \textbf{\bibinfo{volume}{187}},
  \bibinfo{pages}{192} (\bibinfo{year}{1969}).

\bibitem[{\citenamefont{Jobunga}(2018)}]{Jobunga2018}
\bibinfo{author}{\bibfnamefont{E.~O.} \bibnamefont{Jobunga}},
  \bibinfo{journal}{arXiv} p. \bibinfo{pages}{1804.10058}
  (\bibinfo{year}{2018}).

\bibitem[{\citenamefont{Jobunga}(2020{\natexlab{a}})}]{Jobunga2017b}
\bibinfo{author}{\bibfnamefont{E.~O.} \bibnamefont{Jobunga}},
  \bibinfo{journal}{New Horizons in Mathematical Physics}
  \textbf{\bibinfo{volume}{4}}, \bibinfo{pages}{23}
  (\bibinfo{year}{2020}{\natexlab{a}}).

\bibitem[{\citenamefont{Jobunga}(2020{\natexlab{b}})}]{Jobunga2017c}
\bibinfo{author}{\bibfnamefont{E.~O.} \bibnamefont{Jobunga}},
  \bibinfo{journal}{Journal of Advances in Applied Physics}
  \textbf{\bibinfo{volume}{2}}, \bibinfo{pages}{20}
  (\bibinfo{year}{2020}{\natexlab{b}}).

\bibitem[{\citenamefont{Jobunga}(2020{\natexlab{c}})}]{Jobunga2020}
\bibinfo{author}{\bibfnamefont{E.~O.} \bibnamefont{Jobunga}},
  \bibinfo{journal}{Research Gate}
  \textbf{\bibinfo{volume}{RG.2.2.26378.82884/1}}
  (\bibinfo{year}{2020}{\natexlab{c}}).

\bibitem[{\citenamefont{Jobunga and Okeyo}(2020)}]{Jobunga2017e}
\bibinfo{author}{\bibfnamefont{E.~O.} \bibnamefont{Jobunga}} \bibnamefont{and}
  \bibinfo{author}{\bibfnamefont{S.~O.} \bibnamefont{Okeyo}},
  \bibinfo{journal}{Scientific Reports} \textbf{\bibinfo{volume}{10}},
  \bibinfo{pages}{20126} (\bibinfo{year}{2020}).

\bibitem[{\citenamefont{Kotochigova et~al.}(1997)\citenamefont{Kotochigova,
  Levine, Shirley, Stiles, and Clark}}]{Kotochigova1997}
\bibinfo{author}{\bibfnamefont{S.}~\bibnamefont{Kotochigova}},
  \bibinfo{author}{\bibfnamefont{Z.~H.} \bibnamefont{Levine}},
  \bibinfo{author}{\bibfnamefont{E.~L.} \bibnamefont{Shirley}},
  \bibinfo{author}{\bibfnamefont{M.~D.} \bibnamefont{Stiles}},
  \bibnamefont{and} \bibinfo{author}{\bibfnamefont{C.~W.} \bibnamefont{Clark}},
  \bibinfo{journal}{Phys. Rev. A} \textbf{\bibinfo{volume}{55}},
  \bibinfo{pages}{191} (\bibinfo{year}{1997}),
  \urlprefix\url{https://link.aps.org/doi/10.1103/PhysRevA.55.191}.

\bibitem[{\citenamefont{Kraisler et~al.}(2010)\citenamefont{Kraisler, Makov,
  and Kelson}}]{Kraisler2010}
\bibinfo{author}{\bibfnamefont{E.}~\bibnamefont{Kraisler}},
  \bibinfo{author}{\bibfnamefont{G.}~\bibnamefont{Makov}}, \bibnamefont{and}
  \bibinfo{author}{\bibfnamefont{I.}~\bibnamefont{Kelson}},
  \bibinfo{journal}{Phys. Rev. A} \textbf{\bibinfo{volume}{82}},
  \bibinfo{pages}{042516} (\bibinfo{year}{2010}),
  \urlprefix\url{https://link.aps.org/doi/10.1103/PhysRevA.82.042516}.

\bibitem[{\citenamefont{Argaman et~al.}(2013)\citenamefont{Argaman, Makov, and
  Kraisler}}]{Kraisler2013}
\bibinfo{author}{\bibfnamefont{U.}~\bibnamefont{Argaman}},
  \bibinfo{author}{\bibfnamefont{G.}~\bibnamefont{Makov}}, \bibnamefont{and}
  \bibinfo{author}{\bibfnamefont{E.}~\bibnamefont{Kraisler}},
  \bibinfo{journal}{Phys. Rev. A} \textbf{\bibinfo{volume}{88}},
  \bibinfo{pages}{042504} (\bibinfo{year}{2013}),
  \urlprefix\url{https://link.aps.org/doi/10.1103/PhysRevA.88.042504}.

\bibitem[{\citenamefont{Klopper et~al.}(2010)\citenamefont{Klopper, Bachorz,
  Tew, and H\"attig}}]{Klopper2010}
\bibinfo{author}{\bibfnamefont{W.}~\bibnamefont{Klopper}},
  \bibinfo{author}{\bibfnamefont{R.~A.} \bibnamefont{Bachorz}},
  \bibinfo{author}{\bibfnamefont{D.~P.} \bibnamefont{Tew}}, \bibnamefont{and}
  \bibinfo{author}{\bibfnamefont{C.}~\bibnamefont{H\"attig}},
  \bibinfo{journal}{Phys. Rev. A} \textbf{\bibinfo{volume}{81}},
  \bibinfo{pages}{022503} (\bibinfo{year}{2010}),
  \urlprefix\url{https://link.aps.org/doi/10.1103/PhysRevA.81.022503}.

\bibitem[{\citenamefont{Jobunga}(2017)}]{Jobunga2017}
\bibinfo{author}{\bibfnamefont{E.~O.} \bibnamefont{Jobunga}},
  \bibinfo{journal}{arXiv} p. \bibinfo{pages}{1704.02009}
  (\bibinfo{year}{2017}).

\bibitem[{\citenamefont{Jobunga}(2021)}]{Jobunga2021}
\bibinfo{author}{\bibfnamefont{E.~O.} \bibnamefont{Jobunga}},
  \bibinfo{journal}{Research Gate}
  \textbf{\bibinfo{volume}{RG.2.2.17329.30560}} (\bibinfo{year}{2021}).

\bibitem[{\citenamefont{NIST}(2020)}]{NIST}
\bibinfo{author}{\bibnamefont{NIST}},
  \bibinfo{journal}{http://www.physics.nist.gov/PhysRefData/Handbook/Tables}
  (\bibinfo{year}{2020}).

\bibitem[{\citenamefont{Koopmans}(1934)}]{KOOPMANS1934}
\bibinfo{author}{\bibfnamefont{T.}~\bibnamefont{Koopmans}},
  \bibinfo{journal}{Physica} \textbf{\bibinfo{volume}{1}}, \bibinfo{pages}{104
  } (\bibinfo{year}{1934}), ISSN \bibinfo{issn}{0031-8914},
  \urlprefix\url{http://www.sciencedirect.com/science/article/pii/S0031891434900112}.

\end{thebibliography}

\end{document}